\documentclass[useAMS,usenatbib]{mn2e}
\pdfoutput=1
\usepackage{times}
\usepackage{graphicx}
\usepackage{txfonts}
\usepackage{subfigure}
\usepackage{color}
\DeclareGraphicsExtensions{.pdf}

\title{Constraining the Nature of Two Ly$\rm{\balpha}$ Emitters detected by ALMA at z = 4.7}
\author[R.J. Williams et al.]{R.J. Williams$^{1,2}$, J. Wagg$^{1,3,4}$, R. Maiolino$^{1,2}$, C. Foster$^{5,3}$, M. Aravena$^{3,6}$, T. Wiklind$^7$, 
\newauthor C.L. Carilli$^{8,1}$, R.G. McMahon$^9$, D. Riechers$^{10}$, F. Walter$^{11}$ \\ \\
$^1$ Cavendish Laboratory, University of Cambridge, 19 J.J. Thomson Ave., Cambridge, UK\\
$^2$ Kavli Institute for Cosmology, University of Cambridge, Madingley Road, Cambridge, UK\\
$^3$ European Southern Observatory, Alonso de C\'{o}rdova 3107, Casilla 19001, Vitacura Santiago, Chile \\
$^4$ Square Kilometre Array Organisation, Lower Withington, UK\\
$^5$ Australian Astronomical Observatory, PO Box 915, North Ryde, NSW 1670, Australia\\
$^6$ N\'{u}cleo de Astronom\'{\i}a, Facultad de Ingenier\'{\i}a, Universidad Diego Portales, Av. Ej\'{e}rcito 441, Santiago, Chile \\ 
$^7$ Joint ALMA Observatory, Alonso de Cordova 3107, Vitacura, Santiago, Chile\\
$^8$ National Radio Astronomy Observatory, Socorro, NM, USA\\
$^9$ Institute of Astronomy, University of Cambridge, Cambridge, UK\\
$^{10}$ Department of Astronomy, Cornell University, Ithaca, NY, USA\\
$^{11}$ Max-Planck Institute for Astronomy, Heidelberg, Germany }

\begin{document}

	\date{Accepted. Received}
	\pagerange{\pageref{firstpage}--\pageref{lastpage}} \pubyear{2013}

	\maketitle

	\label{firstpage}

%%%%%%%%%%%%%%%%%%%%% ABSTRACT %%%%%%%%%%%%%%%%%%%
\begin{abstract}
We report optical VLT FORS2 spectroscopy of the two Ly$\rm{\alpha}$ emitters (LAEs) companions to the quasi-stellar object (QSO) - sub-millimetre galaxy (SMG) system BRI1202-0725 at $z=4.7$, which have recently been detected in the [CII]158$\rm{\mu m}$ line by the Atacama Large Millimetre/Sub-millimetre Array (ALMA). We detect Ly$\rm{\alpha}$ emission from both sources and so confirm that these Ly$\rm{\alpha}$ emitter candidates are physically associated with the BRI1202-0725 system. We also report the lack of detection of any high ionisation emission lines (N V $\lambda1240$, Si IV $\lambda1396$, C IV $\lambda1549$ and He II $\lambda1640$) and find that these systems are likely not photoionised by the quasar, leaving in situ star formation as the main powering source of these LAEs. We also find that both LAEs have Ly$\rm{\alpha}$ emission much broader ($\sim$$1300$~kms$^{-1}$) than the [CII] emission and broader than most LAEs. In addition, both LAEs have roughly symmetric Ly$\alpha$ profiles implying that both systems are within the HII sphere produced by the quasar. This is the first time that the proximity zone of a quasar is probed by exploiting nearby Ly$\rm{\alpha}$ emitters. We discuss the observational properties of these galaxies in the context of recent galaxy formation models.
\end{abstract}

\begin{keywords}
	galaxies: formation -- galaxies: high-redshift
\end{keywords}

%%%%%%%%%%%%%%%%%%%%% INTRODUCTION %%%%%%%%%%%%%%%%%%%
%%%%%%% FIG %%%%%%%%
\begin{figure*}
	\centering
	\includegraphics[width=0.88\textwidth]{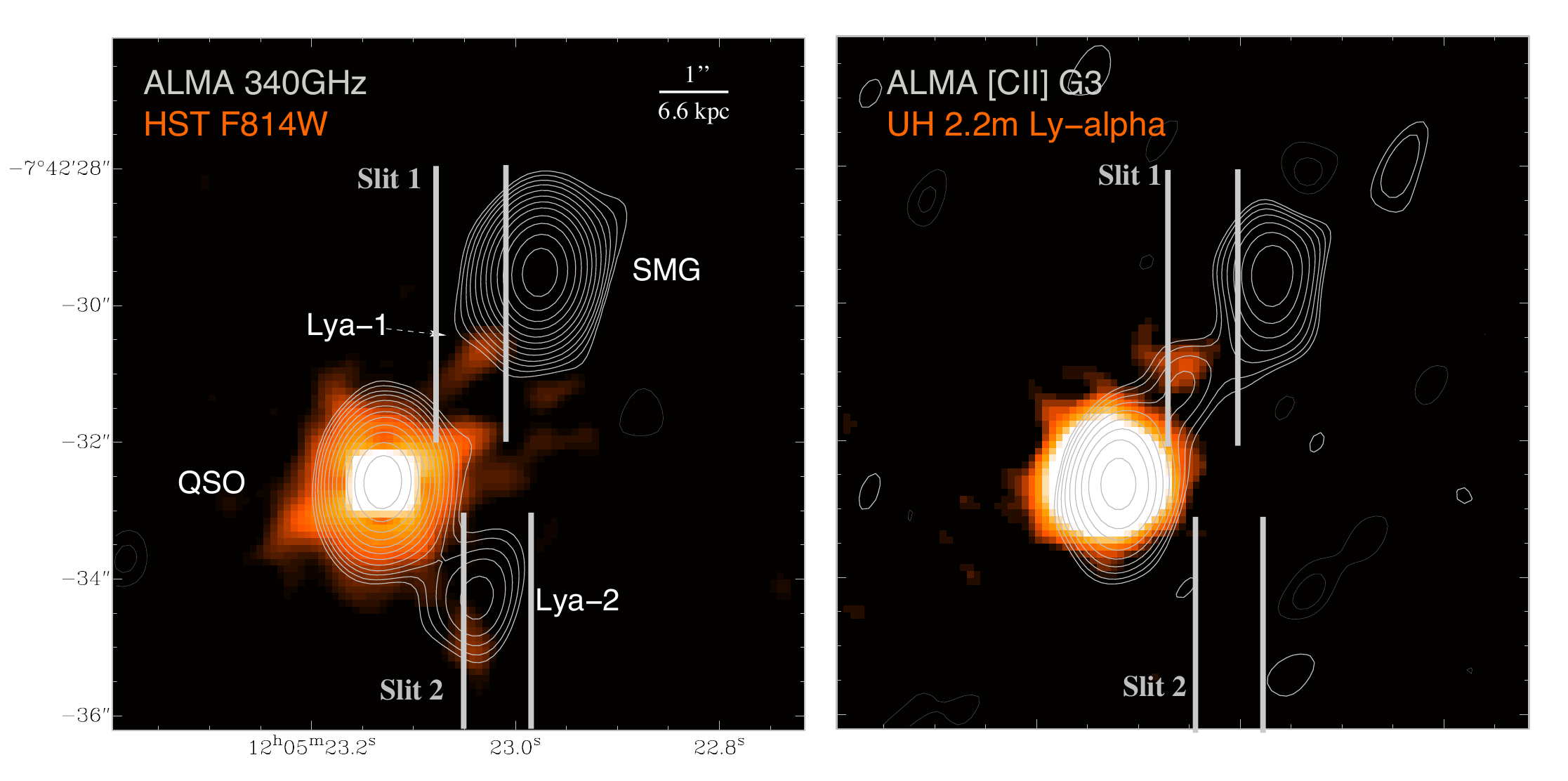}
	\caption{Image of BRI1202-0725 from Carilli et al. (2013) with the location of the FORS2 slits used to study the two Ly$\rm{\alpha}$ sources. Left shows the \textit{HST} F814W image from Hu et al. (1996), while white contours show the sub-millimetre continuum detected by ALMA in Carilli et al (2013). Right shows the narrow band image from Hu et al. (1996) along with the integrated [CII] emission contours detected by ALMA in Carilli et al. (2013).}
	\label{fig:pre_image}
\end{figure*}

\section{Introduction}
Ly$\rm{\alpha}$ emission is a useful probe of star formation in the young Universe and provides an effective means of discovering high-redshift galaxies. This emission can be scattered off of neutral gas and dust in the interstellar medium (ISM) and intergalactic medium (IGM), therefore observations of Ly$\rm{\alpha}$ line emission can provide useful constraints on star-formation in these galaxies. The observed Ly$\rm{\alpha}$ emission from high redshift galaxies is shifted into the optical band and so has the advantage that it can be observed by ground-based telescopes. Galaxies selected for their Ly$\rm{\alpha}$ emission can be surveyed to better understand galaxy formation and evolution \citep[e.g.][]{ellis_08}.

However, it is thought that the star formation traced by optical light only makes up roughly half of the total star formation occurring at high redshifts \citep[e.g][]{devlin_apr09}. The other half is obscured by dust and molecular gas and these galaxies are often undetectable in the optical. This is not true at far infra-red (FIR) through to centimetre (cm) wavelengths, which can be used to trace the dust and molecular gas which fuels early star formation. Since the development of sensitive millimetre (mm)/cm interferometers the number of molecular lines, fine structure lines and continuum detections in high redshift galaxies has increased dramatically \citep[e.g][]{walter_jul03, wang_may10, maiolino_sep12, carilli_jan13}. Even more recently has been the ongoing commissioning of ALMA, which can detect molecular gas to an order of magnitude greater sensitivity than previous facilities.  Therefore through the combination of optical and submm observations, we can now gain a more complete picture of the history of star formation in the young Universe. 

BRI1202-0725 was the first unlensed $z>4$ systems to be detected in submm continuum emission \citep{mcmahon_94, isaak_jul94}. Interferometric observations of molecular CO line emission reveals a QSO and an optically obscured SMG to the North-West \citep{ohta_96, omont_aug96}, both of which are luminous in the FIR and rich in molecular and ionised gas \citep{omont_aug96, riechers_oct06, wagg_jun12, salome_sep12, carilli_feb13, carniani_13}. In addition, narrow band images have revealed extended Ly$\rm{\alpha}$ emission between the QSO and SMG \citep{hu_mar96} and this is also seen in the continuum revealed by \textit{HST} \textit{i}-band images. This emitter is labelled Ly$\rm{\alpha}$-1 by \cite{salome_sep12} and is confirmed spectroscopically to be a Ly$\rm{\alpha}$ emitter at the same redshift as the QSO-SMG system \citep{omont_aug96, ohyama_dec04} and a few 10 kpc from the QSO. A second Ly$\rm{\alpha}$ source is also observed at a similar distance to the South-West of the QSO \citep{hu_mar96} and is labelled Ly$\rm{\alpha}$-2 \citep{salome_sep12}. The redshift of this source is still ambiguous as it has not yet been spectroscopically confirmed. Recent observations of this system using ALMA detects narrow [CII] emission in the vicinity of Ly$\rm{\alpha}$-1, which extends from the QSO to the SMG \citep{carilli_feb13}. Submm continuum emission (at $340$~GHz) is likely associated with Ly$\rm{\alpha}$-2 \citep{wagg_jun12}. The latter is also detected in [CII] by the ALMA observations \citep{wagg_jun12, carilli_feb13}, although at the edge of the ALMA band. \cite{carniani_13} also claim the detection of two additional companions to the QSO and SMG. BRI1202-0725 is clearly a merging group of galaxies, likely on it's way to becoming a giant elliptical in the centre of a dense cluster, therefore represents the best laboratory to date to study early massive galaxy formation.

In this paper we use FORS2 spectroscopic optical observations to study the two Ly$\rm{\alpha}$ emitters in the BRI1202-0725 system, providing a larger bandwidth with extended coverage of other important lines and better sensitivity. We adopt a cosmological model with ($\Omega_{\Lambda}$, $\Omega_{m}$, $h$) = (0.685, 0.315, 0.673) \citep{planck_res13}.

%%%%%%%%%%%%%%%%%%%%% DATA REDUCT. %%%%%%%%%%%%%%%%%%%
\section{Observations and Data reduction}
The FORS2 spectrograph on the VLT was used to perform multi-object spectroscopy on the system of galaxies in the field of BRI1202-0725 on the night of 2013 February 17. The 300I grating centred at $8600$~$\AA$ was used yielding a wavelength coverage between $6000-11000$~\AA.  Slits of $4''$ in length and $1''$ in width were placed on both Ly$\rm{\alpha}$ emitters as in Fig.~\ref{fig:pre_image}, as well as a selection of extended objects identified in the \textit{R}-band pre-image. In total, six sets of observations were taken over one night, each with a thirty minute exposure time.

For the data reduction process we employ part of the FORS2 pipeline as well as self written IDL routines to optimise the data reduction. This is because on initial inspection we find the Ly$\rm{\alpha}$ emission to be extended and so the sky subtraction performed by the pipeline tends to remove source emission.
From the raw scientific frames, we subtract the master bias and perform cosmic ray removal using an IDL script from \cite{vandokkum_nov01}. This removal is based on Laplacian edge detection which relies on the sharpness of the edge of cosmic rays and therefore can be applied to single frames. The \textit{`fors\_calib'} recipe from the pipeline is used to process the calibration exposures associated with each scientific observation thus producing the normalised flat field image, wavelength calibration coefficients and the slit positions on the CCD. These are then used as inputs for the \textit{`fors\_science'} recipe along with the science frame (post cosmic ray removal) to carry out the wavelength calibration. Following this we begin the manual sky subtraction procedure. To do this we take the spectrum from one of the neighbouring slits with no source emission and make  a one-dimensional sky spectrum which we then align with the averaged Ly$\rm{\alpha}$-1 spectrum over a 300 pixel sub-region. Using the best alignment parameters (so as to minimise the residuals between spectra), we re-expand the sky spectrum into two-dimensions and subtract from the Ly$\rm{\alpha}$-1 2D spectrum. We repeat this process for Ly$\rm{\alpha}$-2 by performing a separate alignment for the sky with this spectrum so as to make the optimal subtraction for both cases. Finally, we perform a flux calibration by taking the calibration function obtained by the FORS2 pipeline for a standard star observation and applying it across the spectra.    

We perform the reduction for each of the six observations individually to obtain six spectra, which we then align and stack together. Note that checks on the individual spectra from the reduction reveal that the seeing is similar in each observation ($\sim$$0.6''$), therefore none are excluded from the stack.

%%%%%%% FIG %%%%%%%%
\begin{figure}
	\centering
	\flushleft{a)}
	\includegraphics[width=0.485\textwidth]{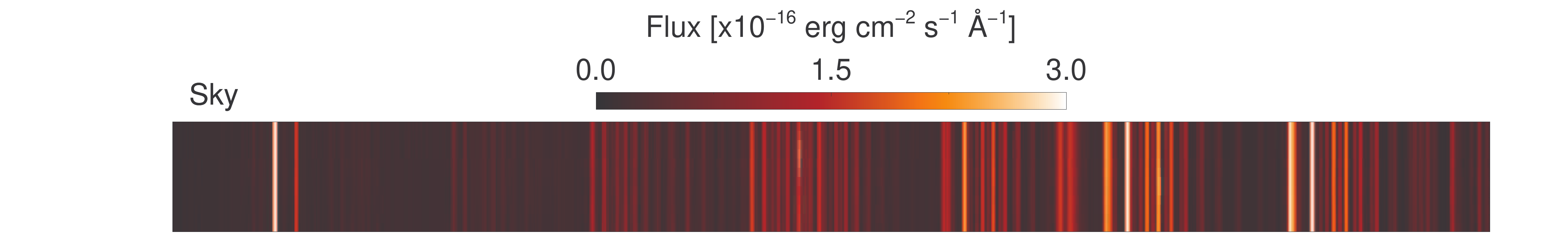}
	\includegraphics[width=0.485\textwidth]{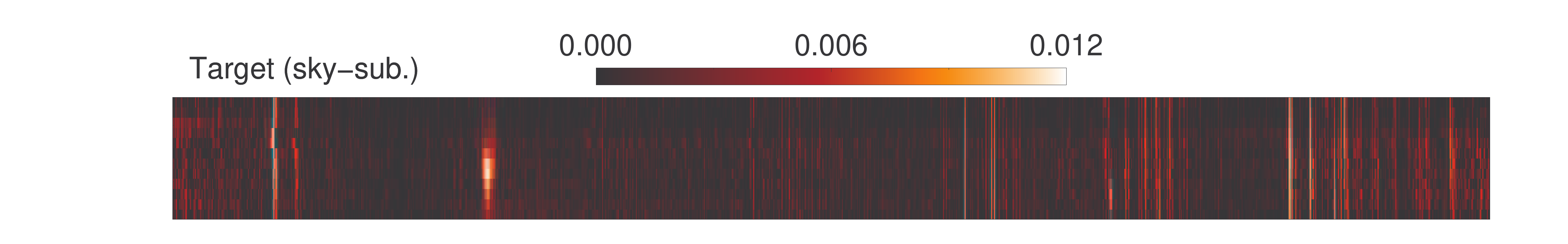}
	\vspace{-1.2cm}
	\includegraphics[width=0.485\textwidth]{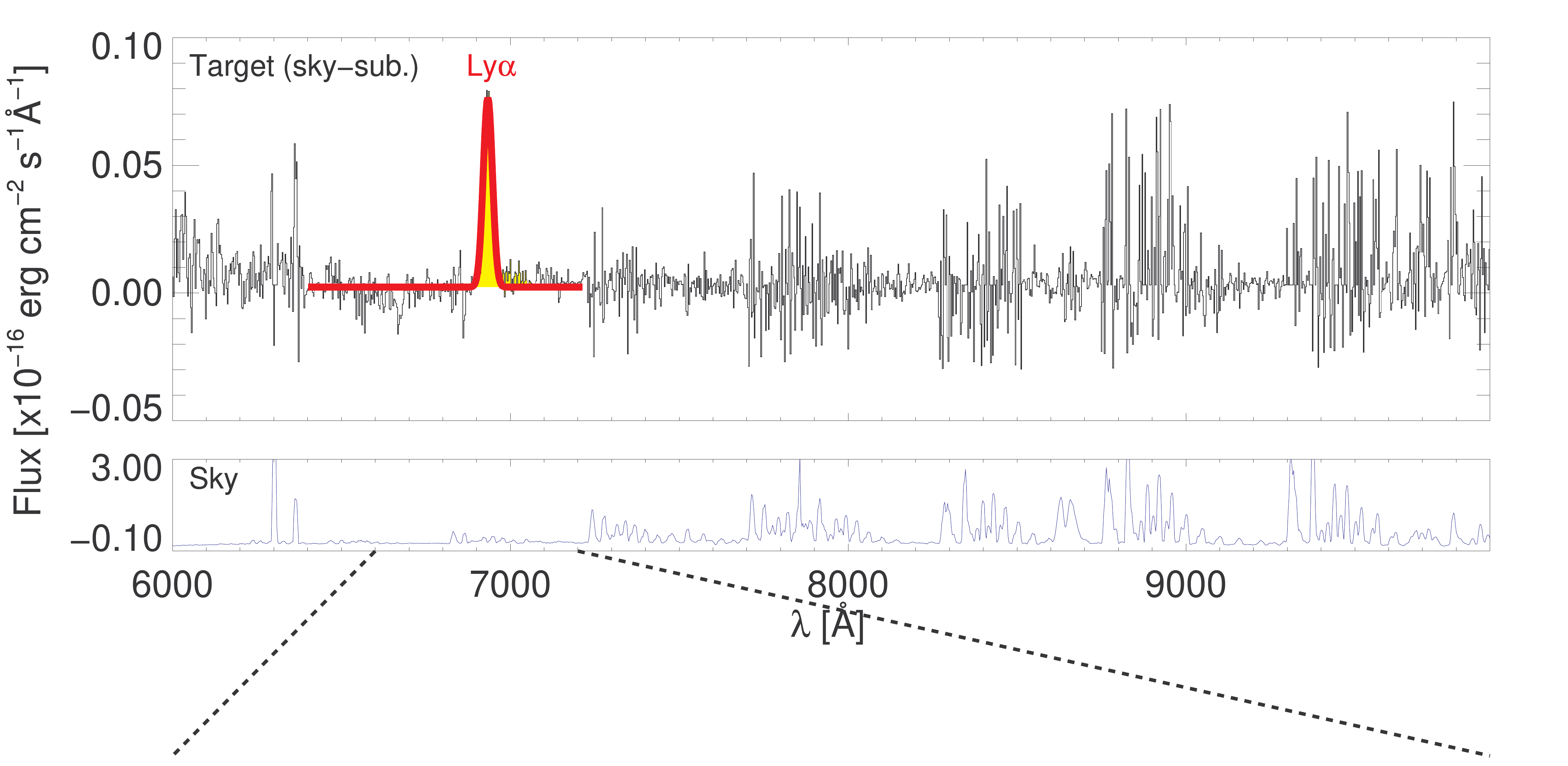}
	\flushleft{b)}
	\includegraphics[width=0.485\textwidth]{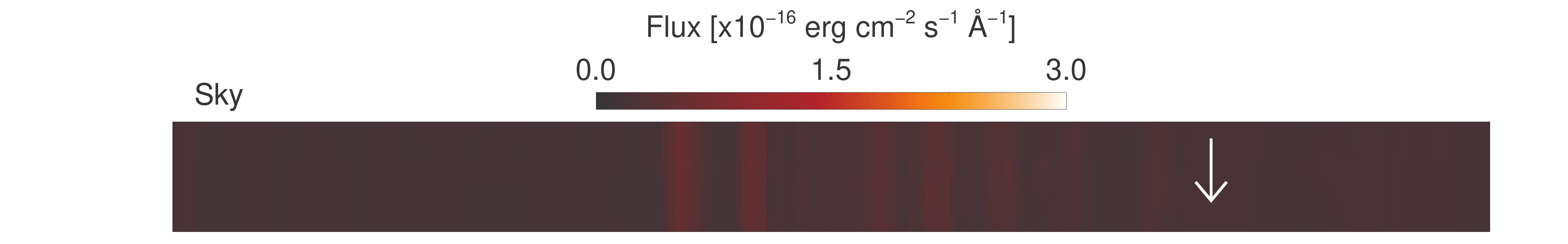}
	\includegraphics[width=0.485\textwidth]{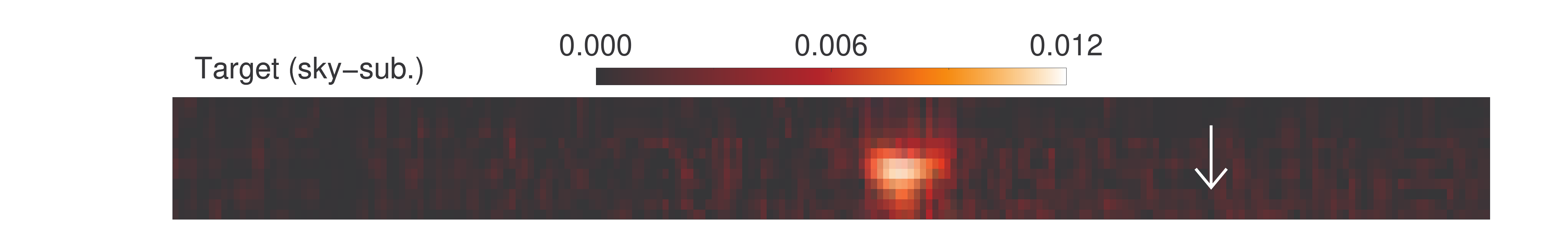}
	\includegraphics[width=0.485\textwidth]{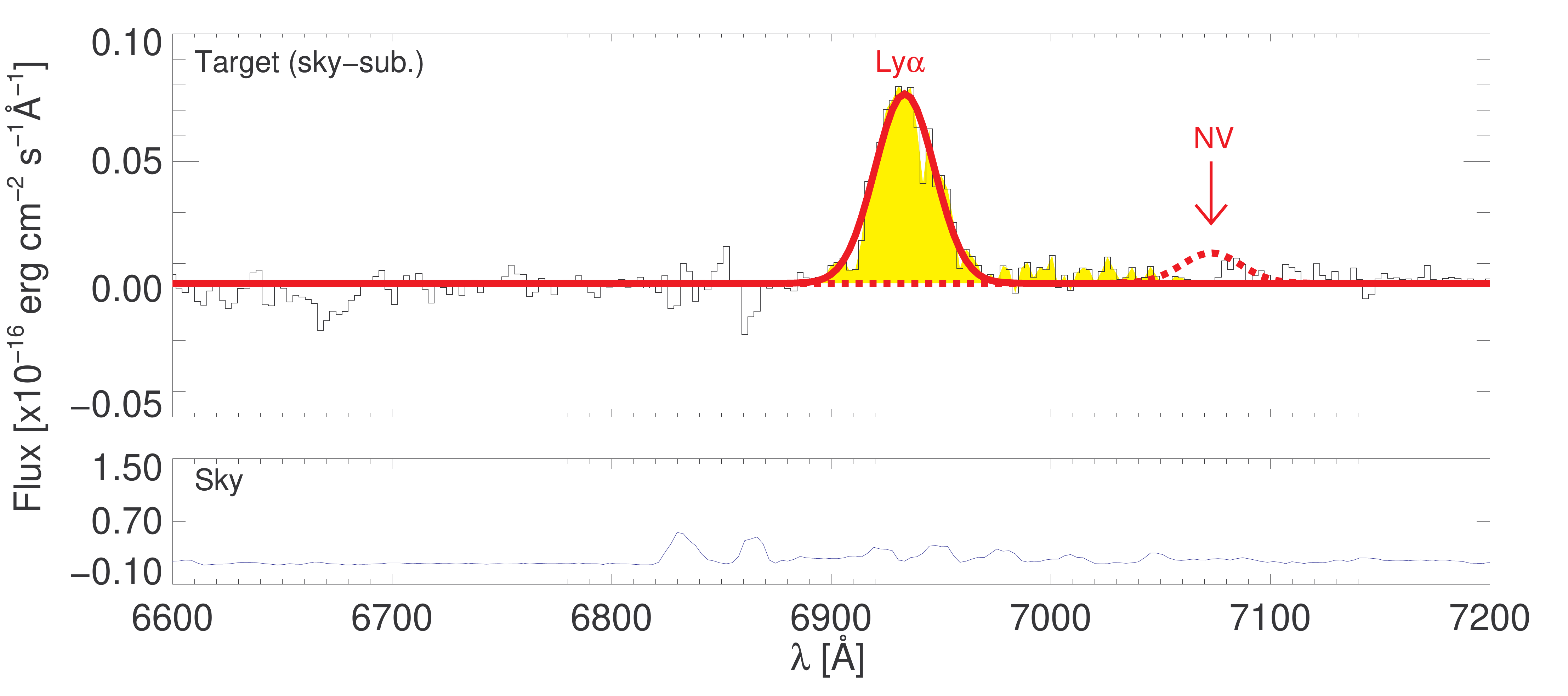}
	\caption{a) The full extracted spectrum obtained from Ly$\rm{\alpha}$-1 and b) a closer view of the Ly$\rm{\alpha}$ emission. The panels from top to bottom show: the 2D sky spectrum, the 2D sky-subtracted galaxy spectrum, the 1D sky-subtracted galaxy spectrum and the 1D sky spectrum showing atmospheric transparency.  The red line shows the Gaussian fit to the emission line. The expected location of N V is indicated by the arrow, with the dashed red line showing the expected emission profile using the average of the line ratios of QSO2s and Sy2s.}
	\label{fig:lya_spec1}
\end{figure}

%%%%%%% TABLE %%%%%%%%
\begin{table}
	\caption{Ly$\rm{\alpha}$ line properties and upper limits on the fluxes of the other lines} 
	\label{tab:properties} 
	\begin{tabular}{ccccc}  
	\hline
	   & Flux [$\times 10^{-16}$  & EW  & Redshift $^{a}$ & FWHM \\
	   & ~erg~cm$^{-2}$~s$^{-1}$] & [$\AA$] & & [kms$^{-1}$]\\
	\hline
	 Ly$\rm{\alpha}$-1 & & & \\
	 Ly$\rm{\alpha}$ & $2.53 \pm 0.08$  & $103 \pm 15$ & $4.703$ & $1381 \pm 124$ \\
	 N V & $<0.05$ & $<2$ & - & -\\
	 Si iV & $<0.04$ & $<2$ & - & -\\
	 C iV & $<0.08$ & $<3$ & - & - \\
	 He II & $<0.05$ & $<3$ & - & -\\
	 	 
	 Ly$\rm{\alpha}-2$ & & & \\
	 Ly$\rm{\alpha}$ & $0.33 \pm 0.06$ & $67 \pm 15$ & $4.698$ & $1225 \pm 257 $ \\
	 N V & $<0.06$ & $<13$ & - & -\\
	 Si iV & $<0.05$ & $<10$ & - & -\\
	 C iV & $<0.08$ & $<15$ & - & - \\
	 He II & $<0.06$ & $<12$ & - & -\\
	\hline
	\end{tabular}\\
	\flushleft{$^{a}$ Error on redshift values $=0.001$} 
\end{table}

%%%%%%%%%%%%%%%%%%%%% RESULTS %%%%%%%%%%%%%%%%%%%

\section{Results}
\begin{figure}
	\centering
	\flushleft{a)}
	\includegraphics[width=0.485\textwidth]{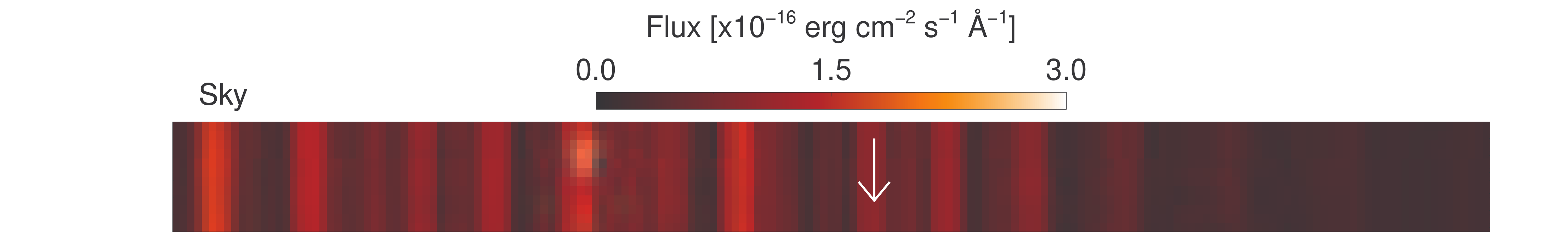}
	\includegraphics[width=0.485\textwidth]{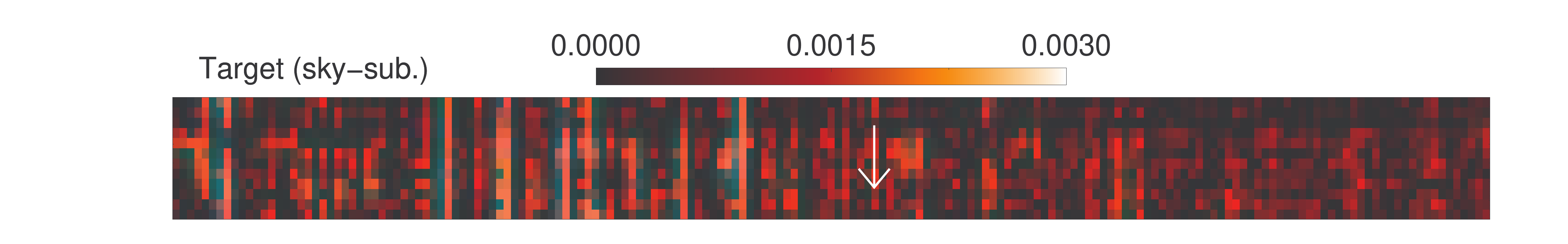}
	\includegraphics[width=0.485\textwidth]{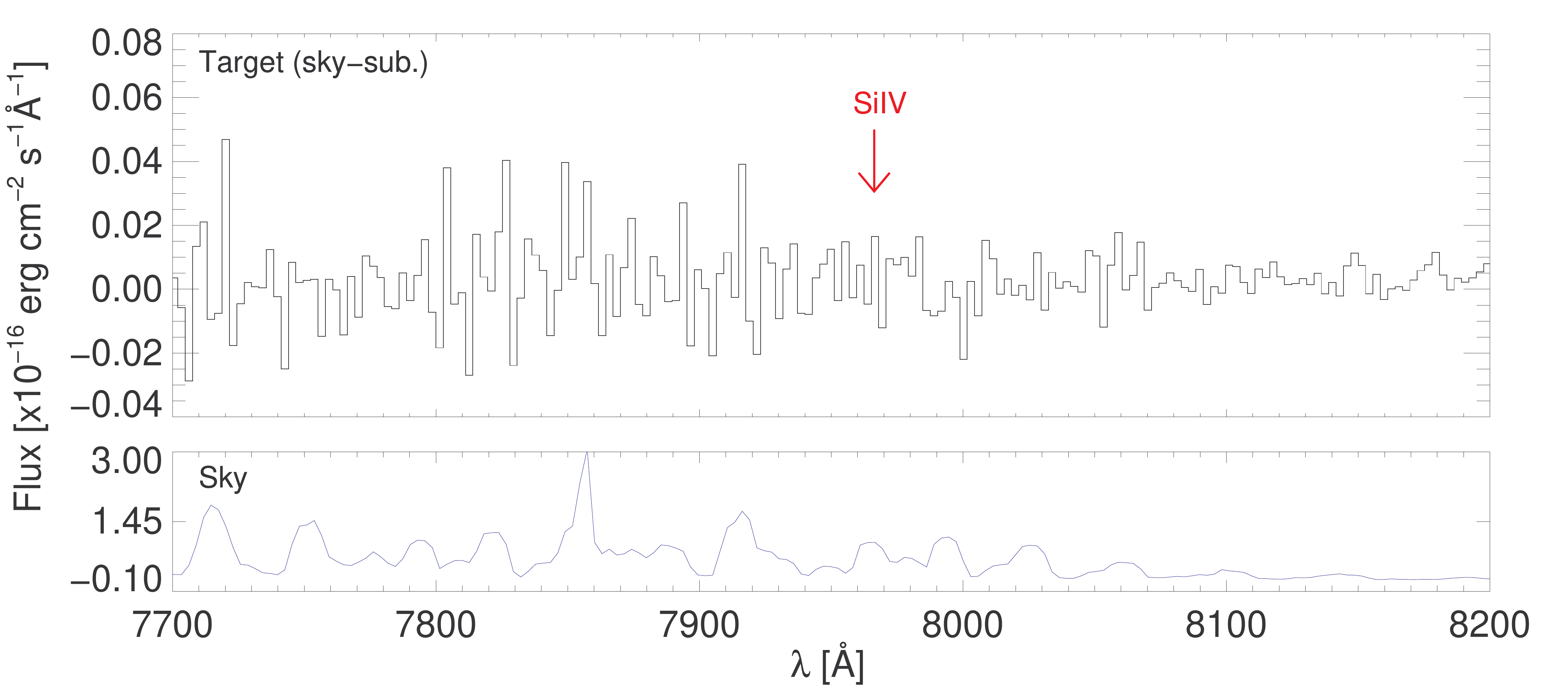}
	\flushleft{b)}
	\includegraphics[width=0.485\textwidth]{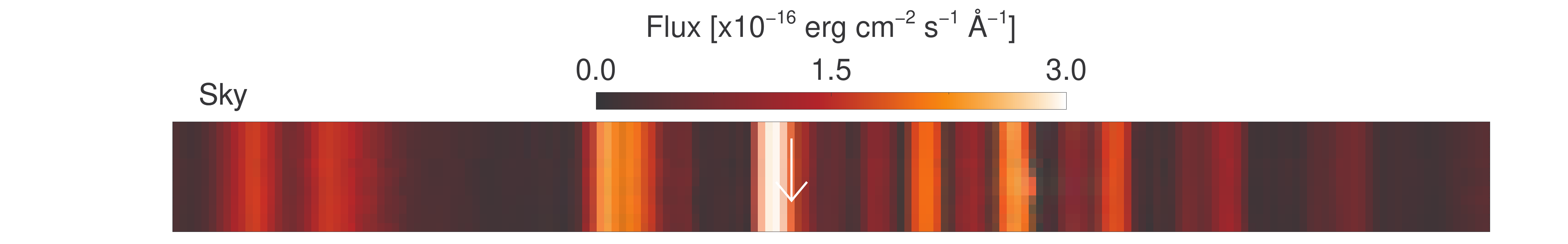}
	\includegraphics[width=0.485\textwidth]{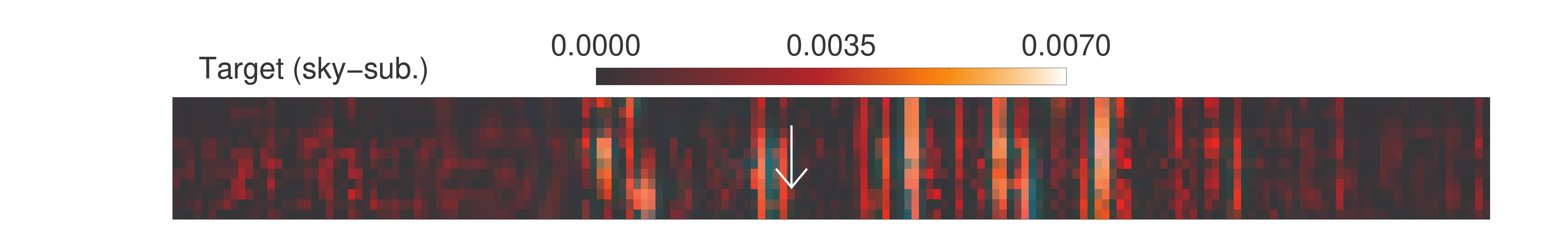}
	\includegraphics[width=0.485\textwidth]{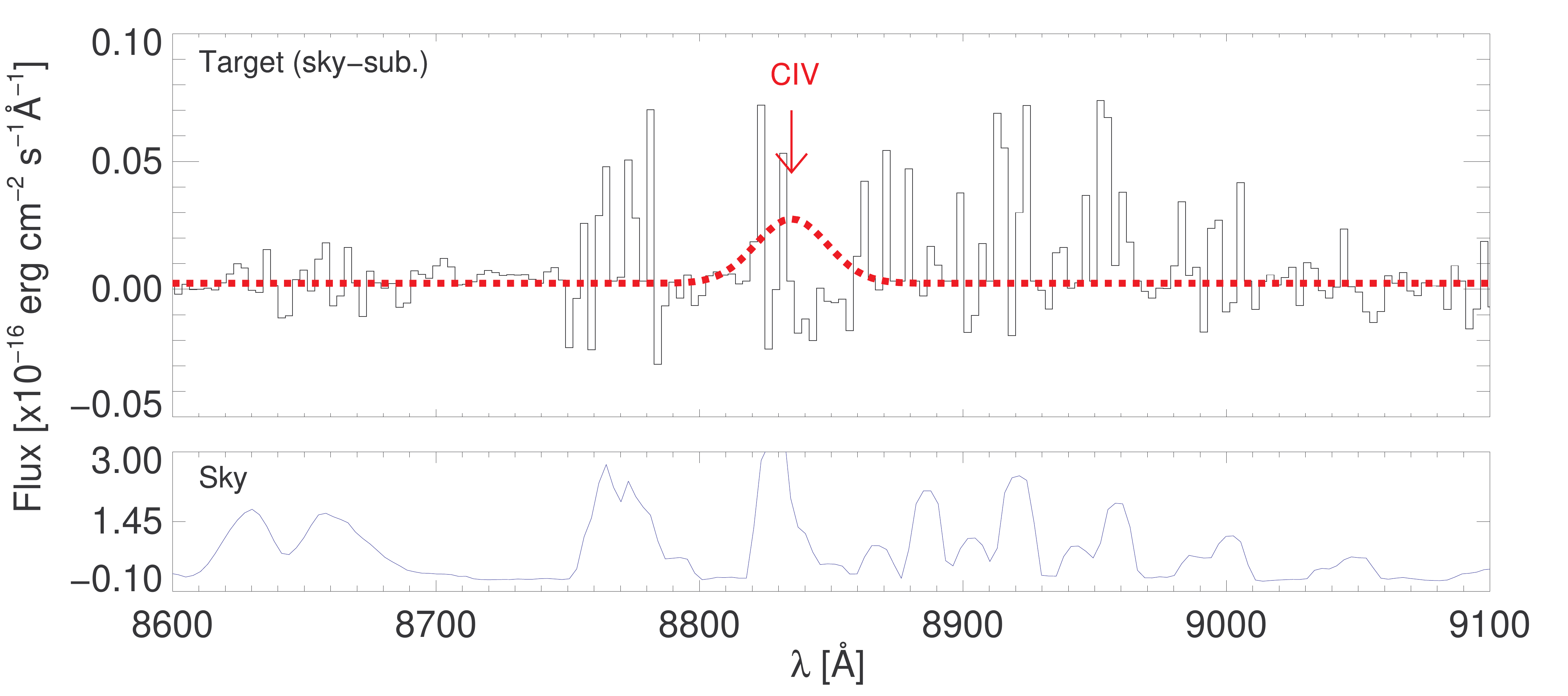}
	\flushleft{c)}
	\includegraphics[width=0.485\textwidth]{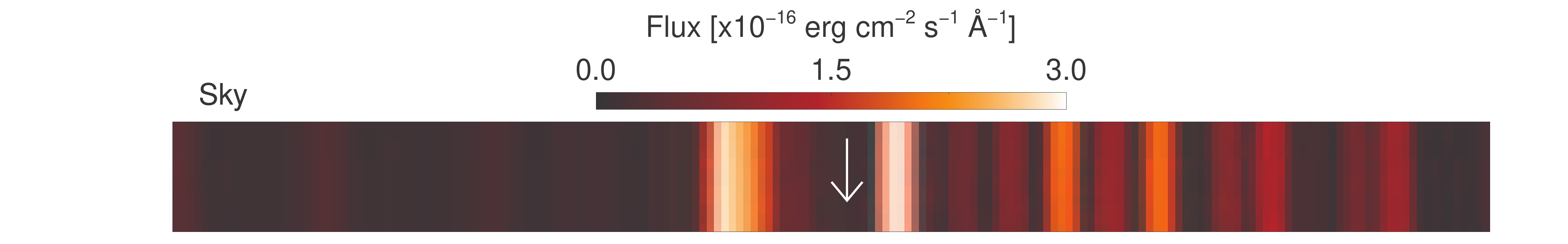}
	\includegraphics[width=0.485\textwidth]{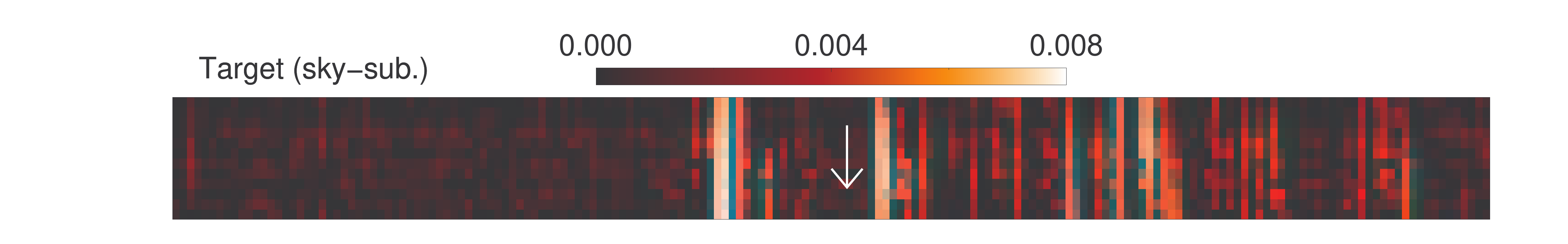}
	\includegraphics[width=0.485\textwidth]{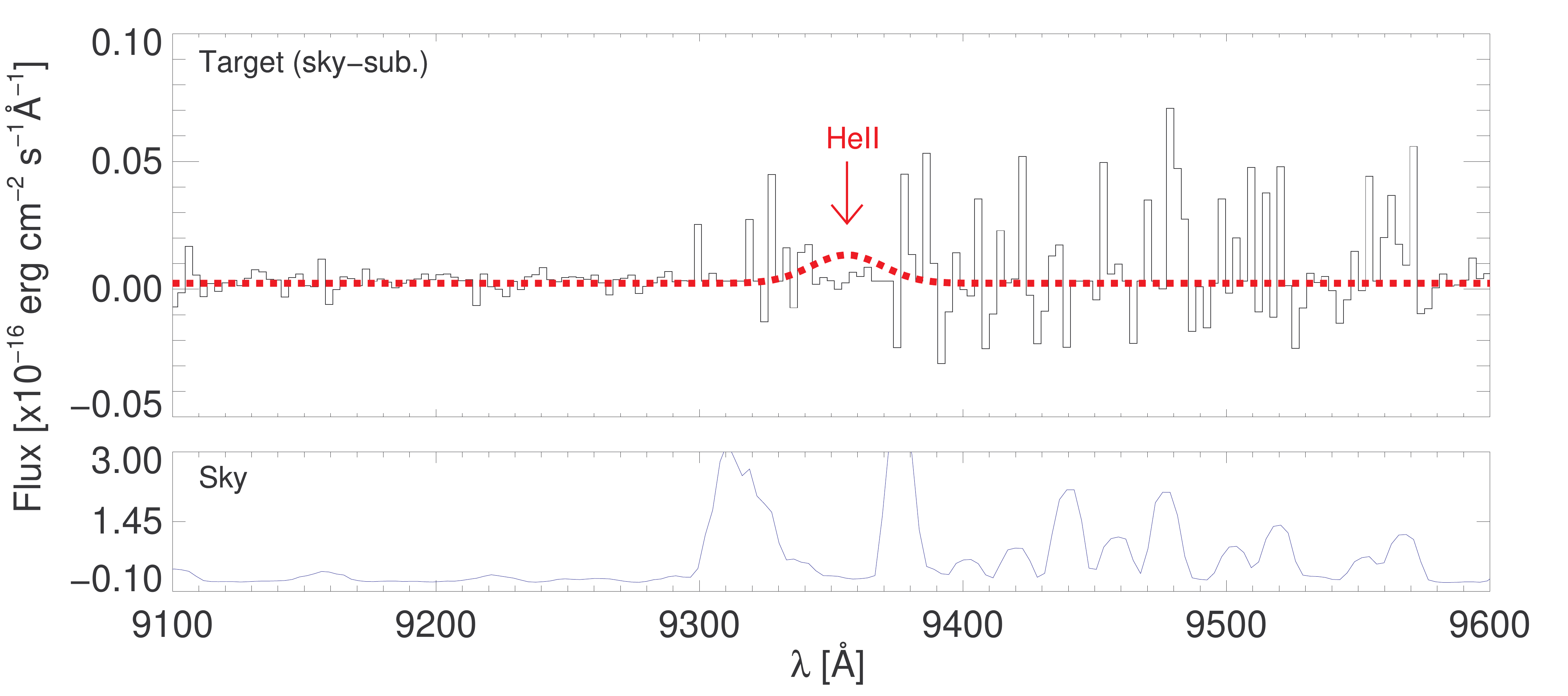}
	\caption{The extracted spectrum of Ly$\rm{\alpha}$-1 showing the expected location of the emission lines a) Si IV, b) C IV and c) He II. The top and bottom panels show the 2D and 1D extracted sky spectrum respectively, as explained in Fig.~\ref{fig:lya_spec1} caption. Again the dashed red line shows the expected emission profile using the average of the line ratios of QSO2s and Sy2s.  }
	\label{fig:non_det}
\end{figure}

In the Ly$\rm{\alpha}$-1 spectrum we detect strong, extended Ly$\rm{\alpha}$ emission with a flux of $2.53 \times 10^{-16}$~erg~cm$^{-2}$~s$^{-1}$. This is $\sim$70\% stronger than that measured by \cite{hu_mar96} with the difference possibly arising due to slit losses, although this remains unclear. The full, reduced spectrum is shown in Fig.~\ref{fig:lya_spec1}. Fitting the Ly$\rm{\alpha}$ line profile with a Gaussian gives the central wavelength of the emission at $\lambda = 6933.46~\rm{\AA}$ which means this source is at redshift $z=4.703$. This is consistent with the spectroscopic redshift determined by \cite{petitjean_apr96}. Fig.~\ref{fig:lya_spec1} also shows that we do not detect the N V emission line, the expected location of which is shown by the arrow. In addition, we do not detect emission lines from Si IV, C IV or He II at the Ly$\rm{\alpha}$ redshift, which we would expect to see. These non-detections are shown in Fig.~\ref{fig:non_det}. Note that in some cases the emission lines are located in regions partly affected by sky lines. However, we are able to compute upper limits on the fluxes for these emission lines (Table~\ref{tab:properties}) and calculate the line ratios with respect to Ly$\rm{\alpha}$ (Table~\ref{tab:ratios}). 

\begin{figure}
	\centering
	\includegraphics[width=0.485\textwidth]{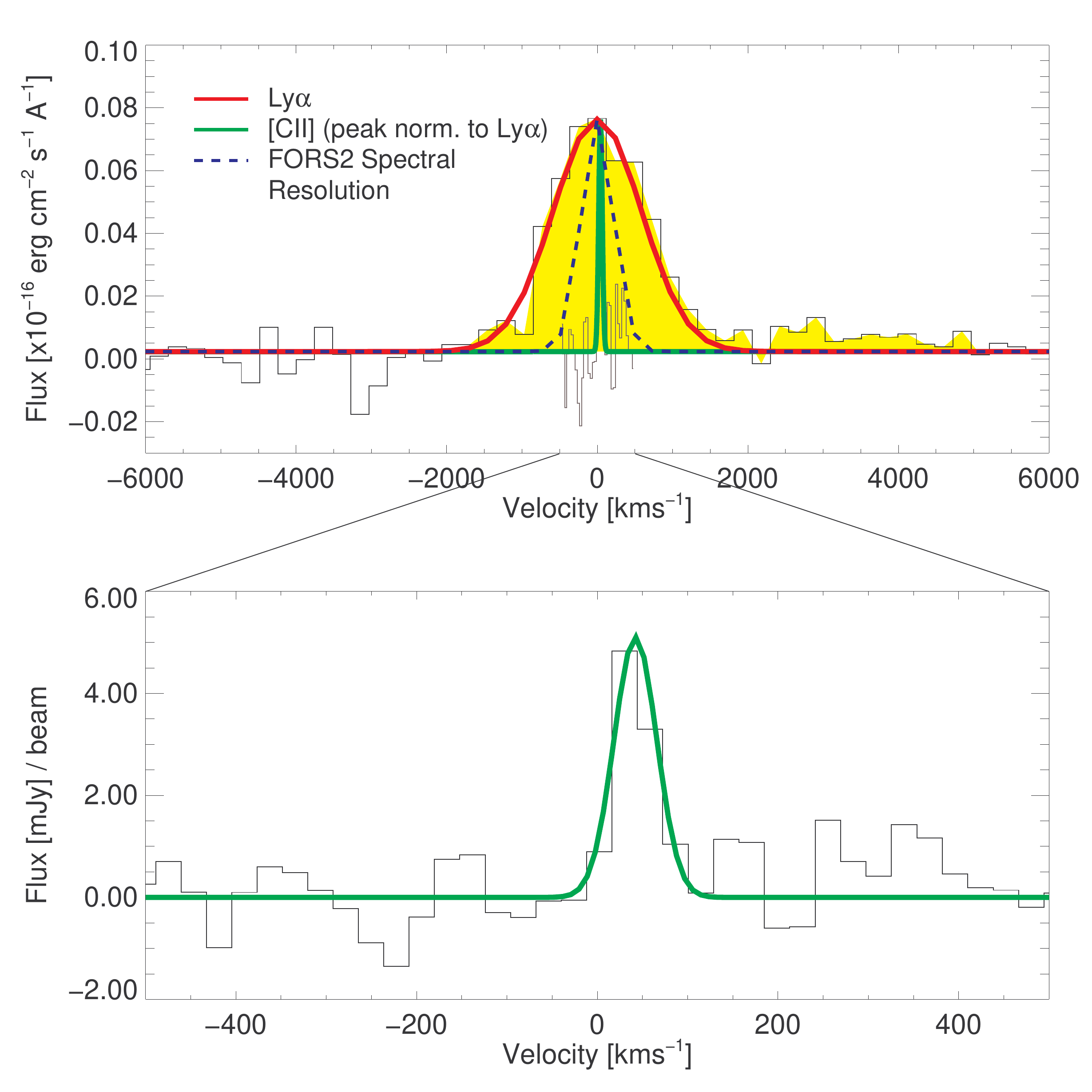}
	\caption{Comparison of the Ly$\rm{\alpha}$ and [CII] profiles for Ly$\rm{\alpha}$-1. Top: spectrum of the Ly$\rm{\alpha}$ line (the red line shows the Gaussian fit), compared with the FORS2 instrumental resolution (blue dashed line), and with the [CII] profile (the green line shows the associated Gaussian fit, with the peak normalised to the Ly$\rm{\alpha}$ peak) taken from \citet{carilli_feb13}. Bottom: expanded view of the [CII] spectrum. } 
	\label{fig:compare}
\end{figure}

For the Ly$\rm{\alpha}$-1 galaxy, we compare our Ly$\rm{\alpha}$ profile to that of the [CII] emission from \cite{carilli_feb13}, as shown in Fig.~\ref{fig:compare}. We find the two profiles to be slightly shifted with respect to each other by $\Delta v = 49$~kms$^{-1}$ and, more importantly, the Ly$\rm{\alpha}$ line is much broader ($\sim1400$~kms$^{-1}$) than the [CII] line ($56$~kms$^{-1}$). Interestingly, we tentatively detect a high velocity tail of Ly$\rm{\alpha}$ at the $2\sigma$ level extending to $5000$~kms$^{-1}$, which can be seen more clearly in Fig.~\ref{fig:compare}. Note that the spectral region covered by Ly$\rm{\alpha}$ is relatively clean from strong OH sky lines.

\begin{figure}
	\centering
	\includegraphics[width=0.485\textwidth]{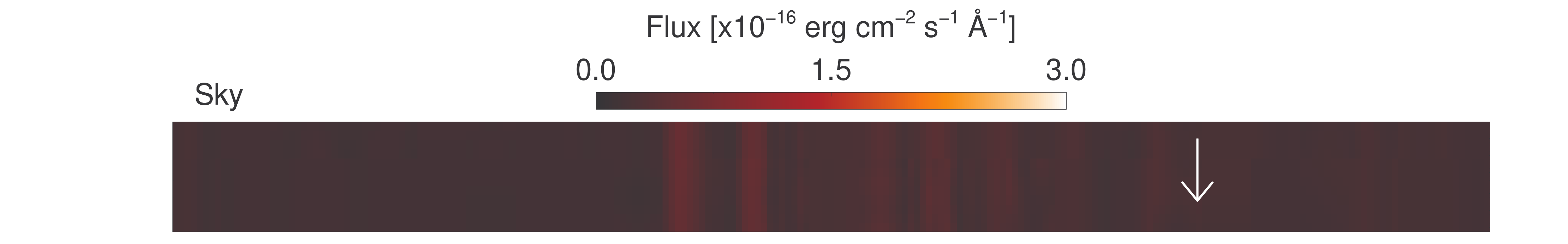}
	\includegraphics[width=0.485\textwidth]{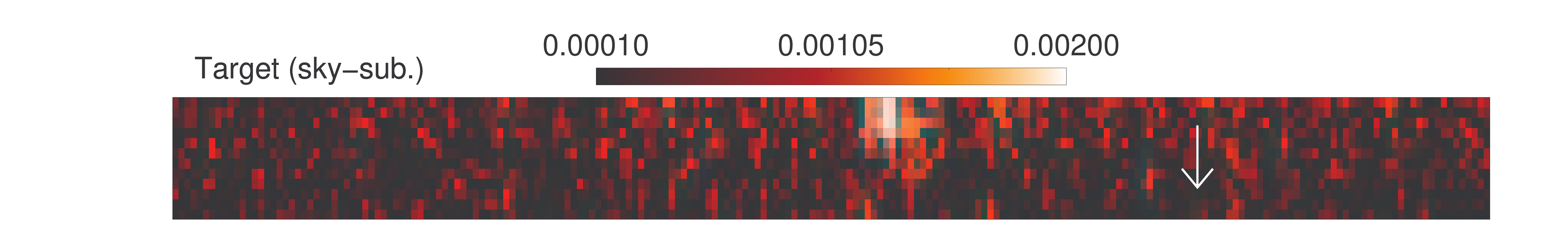}
	\includegraphics[width=0.485\textwidth]{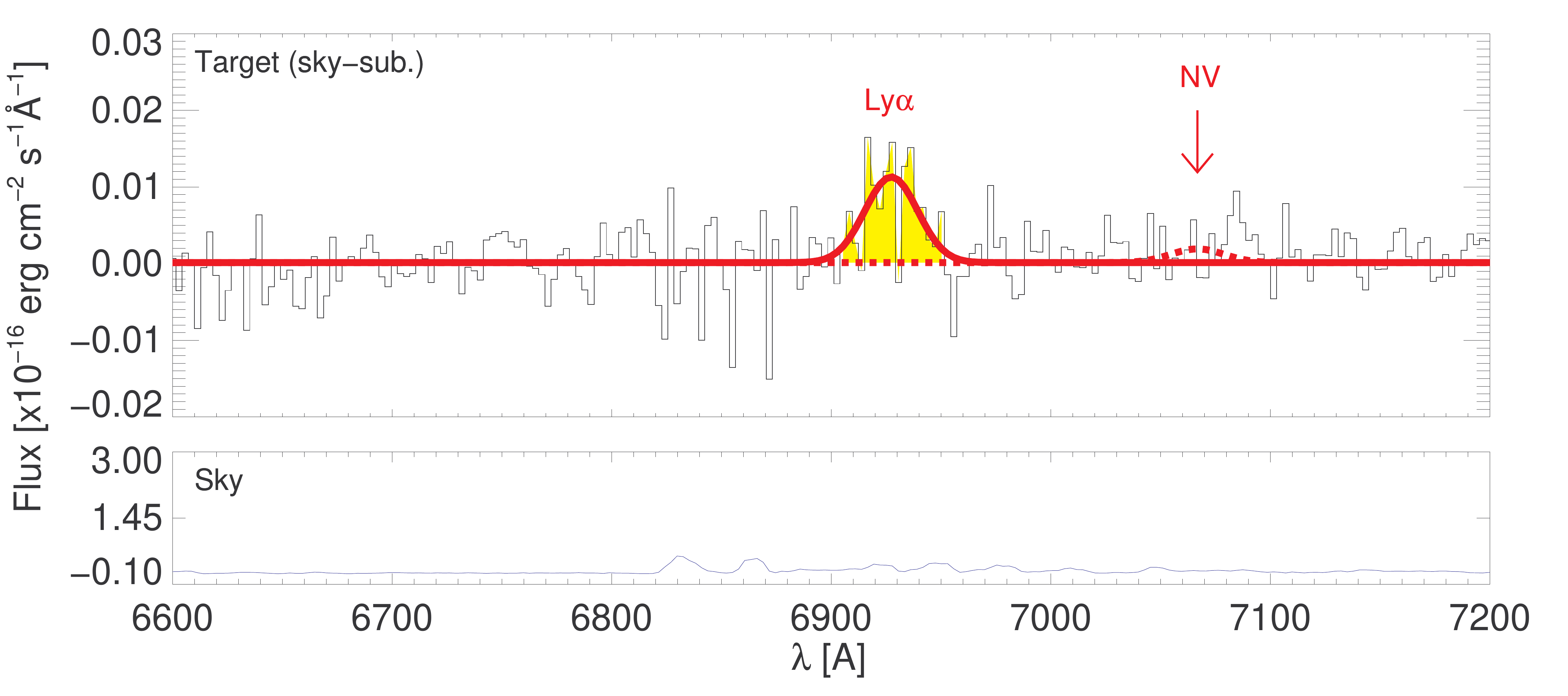}
	\caption{Ly$\rm{\alpha}$ emission from extracted spectrum for Ly$\rm{\alpha}$-2 . The top and bottom panels show the 2D and 1D extracted sky spectrum respectively, as explained in Fig.~\ref{fig:lya_spec1} caption. The red line shows the Gaussian fit to the emission line. And again, the expected location of N V is indicated by the arrow, with the dashed red line showing the expected emission profile using the average of the line ratios of QSO2s and Sy2s.}
	\label{fig:lya_spec2}
\end{figure}

In addition, we detect Ly$\rm{\alpha}$ emission from the Ly$\rm{\alpha}$-2 galaxy, though at a much lower significance. This is shown in Fig.~\ref{fig:lya_spec2} with a flux of $0.33 \times 10^{-16}$~erg~cm$^{-2}$~s$^{-1}$. The Gaussian fitting gives a central wavelength of $\lambda = 6927.32~\rm{\AA}$ corresponding to a redshift $z=4.698$. This is also slightly shifted with respect to the [CII] emission detected from this galaxy \citep{carilli_feb13} by $\Delta v = 45$~kms$^{-1}$ and the Ly$\rm{\alpha}$ line is also much broader than the corresponding [CII] line. Again, we do not detect emission lines from N V, Si IV, C IV or He II, although in this case the upper limits on the line fluxes relative to Ly$\rm{\alpha}$ are much less stringent than in Ly$\rm{\alpha}$-1. A summary of the Ly$\rm{\alpha}$ emission properties and non-detections for both sources are shown in Table~\ref{tab:properties}.

%%%%%%% TABLE %%%%%%%%
\begin{table}
	\caption{Emission line ratio upper limits for Ly$\rm{\alpha}$-1 compared to the average ratios for the \citet{nagao_mar06} sample of QSO2s and Sy2s.} 
	\label{tab:ratios} 
	\begin{tabular}{ccccc}  
	\hline
	  &N V/ Ly$\rm{\alpha}$ & Si IV/ Ly$\rm{\alpha}$ & C IV/ Ly$\rm{\alpha}$ & He II/ Ly$\rm{\alpha}$\\
	\hline
	 Ly$\rm{\alpha}$-1 & $<0.019$ & $<0.017$ & $<0.033$ & $<0.018$ \\
	Nagao QSO2s & $0.163$ & - & $0.337$ & $0.149$\\
	Dispersion & $0.113$ & - & $0.167$ & $0.074$\\
	Nagao Sy2s & $0.083$ & - &$0.284$ & $0.159$\\
	Dispersion & $0.106$ & - & $0.212$ & $0.072$\\
	\hline
	\end{tabular}
\end{table}

%%%%%%% FIG %%%%%%%%

%%%%%%% FIG %%%%%%%%

%%%%%%% FIG %%%%%%%%

%%%%%%%%%%%%%%%%%%%%% DISCUSSION %%%%%%%%%%%%%%%%%%%

\section{Discussion}
\subsection{Ly$\rm{\balpha}$-1}
Ly$\rm{\alpha}$ is the only line detected in this object, all other high ionisation lines that would be expected in the case of clouds photoionised by a quasar are absent. In Table~\ref{tab:ratios} we list the average ratios (and dispersion) of high-ionisation emission lines observed in high-z type 2 QSOs (QSO2s) and Seyfert 2 galaxies (Sy2s), relative to their Ly$\rm{\alpha}$ emission; these ratios are clearly much higher than the stringent upper limits obtained from our spectrum, listed in the same table. To better illustrate the expected flux of such high ionisation lines in the case of photoionisation by a quasar in Fig.~\ref{fig:lya_spec1}-\ref{fig:non_det} we show with a dashed profile the expected intensity of the various lines in the case of the line ratios (relative to Ly$\rm{\alpha}$) typical of type 2 active galactic nuclei (AGNs). Of course, if Ly$\rm{\alpha}$ at this redshift suffers from some IGM attenuation, then the expected fluxes of the high ionisation lines should be even higher, so our expected fluxes are conservative. 
The lack of detection of C IV is particularly important given that this line is very strong in AGN photoionised regions. In addition, the non-detection of He II is also significant. Indeed, if Ly$\rm{\alpha}$-1 was a system with very low metallicity, the high ionisation lines would not be detectable even in the case of AGN photoionisation, however He II should be detectable regardless of metallicity. By comparing our He II/ Ly$\rm{\alpha}$ upper limit with the value typically observed in type 2 AGNs (see Table~\ref{tab:ratios}) we estimate that the fraction of Ly$\rm{\alpha}$ produced as a consequence of ionisation by the QSO must be less than $10\%$. As a consequence, the main likely explanation for both Ly$\rm{\alpha}$ and [CII] emission in this source is that they are powered by in situ star formation. 

This source, together with Ly$\rm{\alpha}$-2, are the first $z>3$ Ly$\rm{\alpha}$ emitters detected in the [CII]158$\rm{\mu m}$ line. 
 The [CII] emission \citep{carilli_feb13} together with the \cite{sargsyan_aug12} conversion factor, indicate a SFR$\sim$$19~$M$_{\odot}$yr$^{-1}$ for this system, which is nicely consistent with the SFR inferred from the rest-frame UV emission $\rm{SFR}_{\rm{UV}}$$\sim$$13~\rm{M_{\odot}~yr}^{-1}$, \citep{ohyama_dec04}. We note that the size of the Ly$\rm{\alpha}$ emission region measured in the FORS2 slit is $\sim2.2''$, while the $340$~GHz ALMA beam is $\sim1''$. Since the [CII] emission is unresolved, this would imply that the [CII] emitting region is smaller than the ionised region. This would be consistent with recent simulations of the ISM in young star forming systems at high redshift \citep[e.g.][]{vallini_aug13}, which indeed expect strongly different distributions between ionised gas (traced by Ly$\rm{\alpha}$) and [CII] emission. However, it is also possible that there is a more extended [CII] emission line component below the sensitivity of current observations.

The EW(Ly$\rm{\alpha}$) is very high, similar to other high-z Ly$\rm{\alpha}$ emitters. This indicates that this is either a young system (on the order of a few Myrs) or is an older, very metal poor system \citep{schaerer_jan03}, though the latter is unlikely due to the detection of [CII].

The most striking feature of this system is the large width of the Ly$\rm{\alpha}$ emission ($1381$~kms$^{-1}$), which is uncommon among any other Ly$\rm{\alpha}$ emitters and is accompanied by an even higher velocity tail ($5000$~kms$^{-1}$). 
The drastically different kinematics between Ly$\rm{\alpha}$ and [CII] in Ly$\rm{\alpha}$-1 may trace an extremely inhomogeneous ISM and star formation environment in these early, young star forming systems. This would again be in line with the strongly different distributions (hence different kinematics) between ionised gas and [CII] emission in high-z galaxies expected by recent models \citep{vallini_aug13}, as a consequence of the complex ISM stratification and distribution of the UV radiation field. 

However, the large velocity range traced by Ly$\rm{\alpha}$, at odds with the other LAE at similar redshift, remains puzzling and must be associated with the specific environment this system is residing in. More specifically, the observed high velocities may indicate that the system is subject to strong tidal shearing by either the QSO host, or by the SMG, or possibly both. Within this tidally stretched system, compact clumps may be less affected by tidal forces and these may be traced by the narrow [CII]. However, the distance of these systems from the QSO and the SMG (at least a few $10$~kpc, by accounting also for the redshift difference), suggests that the tidal scenario is unlikely.  \cite{ohyama_dec04} also suggest the possibility of superwinds originating from Ly$\rm{\alpha}$-1 to explain the elongated morphology, however the width of Ly$\rm{\alpha}$ emission is possibly too large for this scenario when compared to the theoretical velocities found for starburst driven winds \citep[e.g.][]{martin_mar05, thacker_dec06}.

Another interesting possibility is that this could be tracing the ``quasar positive feedback"  predicted by some recent models \citep[e.g.][] {ishibashi_dec12, zubovas_jun13, silk_aug13}. According to these models, gas ejected out of the quasar host galaxy may condense and start forming stars in the outflow. The very broad velocity profile (including the very high velocity tail) of Ly$\rm{\alpha}$ in this star forming system as well as the young age, $\sim$$10$~Myr \citep{ohyama_dec04} inferred from the SED fitting, would fit nicely in this scenario. 

In alternative, both systems could be companions formed independently of the QSO, but their outer regions are blown away to high velocities by the QSO radiation pressure. However, in this case one would expect that the high velocity ionised gas should show some signatures
of photoionisation by the QSO, i.e. high ionisation lines and He II in particular, which are not detected.

It is also of interest that there is a lack of asymmetry in both the two Ly$\rm{\alpha}$ profiles, in contrast with any other LAE at such high redshift, whose Ly$\rm{\alpha}$ blue shoulder is heavily absorbed by the intervening neutral IGM. This implies that no neutral IGM is present in the vicinity of either system further implying that both systems are within the HII sphere produced by quasar. This is the very first time that the HII region surrounding a high-z quasar is probed with such a method. Considering the distance between the QSO and the LAEs we can place a lower limit on the radius of the HII sphere at greater than $15$~kpc.

\subsection{Ly$\rm{\balpha}$-2}
In the case of Ly$\rm{\alpha}$-2, we provide the first confirmation that this galaxy is a Ly$\rm{\alpha}$ emitter at roughly the same redshift as the quasar and so part of the same group, which had still remained unclear previous to these results. The Ly$\rm{\alpha}$ emission is much weaker than in Ly$\rm{\alpha}$-1, which is likely a consequence of dust obscuration. Indeed this system also shows strong far-IR emission, as inferred by the ALMA continuum data \citep{wagg_jun12, carilli_feb13} also implying that this system is likely more metal rich than Ly$\rm{\alpha}$-1. The constraints on the high ionisation lines are much looser and in principle we cannot exclude the presence of N V, Si IV, C IV or He II at a level, relative to Ly$\rm{\alpha}$, compatible with AGN photoionisation. However, we find that the strong far-IR emission ($L_{\rm{FIR}}=1.7 \times 10^{12} L_{\odot}$) cannot be explained in terms of dust heating by the QSO at the distance of Ly$\rm{\alpha}$-2 (even complete absorption of the quasar radiation would yield to a dust re-emitted far-IR luminosity less than $2.3 \times 10^{11} L_{\odot}$) therefore is not consistent with solely dust heating from the quasar, hence also requiring in situ star formation. Again the width of Ly$\rm{\alpha}$ is significantly larger than [CII], although in this case the difference is less extreme, probably because [CII] is at the edge of the ALMA band and therefore it is more difficult to measure its profile. However, the possible interpretation given for the nature of Ly$\rm{\alpha}$-1 (tidal shearing and result of quasar feedback), can certainly apply to this object as well. The Ly$\rm{\alpha}$ profile of this object is also symmetric, indicating that also this object is located within the HII region produced by the quasar.

%%%%%%%%%%%%%%%%%%%%% CONCLUSION %%%%%%%%%%%%%%%%%%%

\section{Conclusion}
In summary, we have presented the results from our FORS2 spectroscopic observations of the two Ly$\rm{\alpha}$ emitters in the BRI1202-0725 system.
\begin{enumerate}
	\item We detect Ly$\rm{\alpha}$ emission from Ly$\rm{\alpha}$-1 and Ly$\rm{\alpha}$-2, which spectroscopically confirms that these are both Ly$\rm{\alpha}$ emitters physically associated with the BRI1202-0725 system \citep{ohyama_dec04}.
	\item The observed Ly$\rm{\alpha}$ emission must be produced by in situ star formation (i.e. no quasar photoionisation/ fluorescence), which is mostly a consequence of the lack of He II emission.
	\item The very large velocity widths of the Ly$\rm{\alpha}$ emission profiles ($\sim$$1100-1400$~kms$^{-1}$, with a possible tentative tail extending to $5000$~kms$^{-1}$) are peculiar for these two star forming Ly$\rm{\alpha}$ systems, which may indicate that these are sites of induced star formation in the quasar outflow as proposed by several recent models of positive feedback.
	\item The lack of asymmetry in the two Ly$\rm{\alpha}$ profiles implies that no neutral IGM is present in the vicinity of either system, implying that both systems are within the HII sphere produced by the quasar. Therefore we place a lower limit on the radius of the HII sphere at greater than $15$~kpc. 
\end{enumerate}

\subsection{Acknowledgements}
We are grateful to Andrea Ferrara for useful discussions.
This work was co-funded under the Marie Curie Actions of the European Commission (FP7-COFUND). Based on observations made with ESO Telescopes at the La Silla Paranal Observatory under programme ID 289.A-5021. This paper makes use of the following ALMA data: ADS/JAO.ALMA\#2011.0.00006.SV. ALMA is a partnership of ESO (representing its member states), NSF (USA) and NINS (Japan), together with NRC (Canada) and NSC and ASIAA 
(Taiwan), in cooperation with the Republic of Chile. The Joint ALMA 
Observatory is operated by ESO, AUI/NRAO and NAOJ. The National Radio Astronomy Observatory is a facility of the National Science Foundation operated under cooperative agreement by Associated Universities, Inc.

%%%%%%%%%%%%%%%%%%%%% REFERENCES %%%%%%%%%%%%%%%%%%%

\bibliographystyle{mn2e}
\setlength{\labelwidth}{0pt}
\bibliography{myrefs}

\label{lastpage}
\end{document}